\documentclass[12pt]{article}
\usepackage{amsmath}
\usepackage{amssymb}
\usepackage{epsfig}
\usepackage{cite}

\def\beq{\begin{equation}}
\def\eeq{\end{equation}}
\def\half{{\textstyle {\frac12}}}

\def\threehalf{{\textstyle {\frac32}}} 
  
 \def\quart{{\textstyle {\frac14}}}
\def\M#1{{{\rm\bf M}}\left[#1\right]}
\def\Min#1{{{\rm\bf M}}^{-1}\!\left[#1\right]}
\def\ca{{C^{}_A}}
\def\cf{{C^{}_F}}

\def\sgn#1{{\mathop{\rm sgn} #1}}

\def\abs#1{\left|#1\right|}

\def\beeq{\begin{eqnarray}}
\def\eeeq{\end{eqnarray}}
\def\bsub{\begin{subequations}}
\def\esub{\end{subequations}}

\def\cO#1{{\cal{O}}\!\left(#1\right)}
\def\guni{\gamma_{\mbox{\scriptsize uni}}}
\def\alphys{{a}_{\rm ph}}
\def\alphaphys{{\alpha}_{\rm ph}}
\def\ee{e^+e^-}
\def \as{\relax\ifmmode\alpha\else{$\alpha${ }}\fi}
\def \al{\relax\ifmmode\alpha\else{$\alpha${ }}\fi}

\def\hS#1{\hat{S}_{#1}}
\def\hhS#1{\underline{S_{#1}}}
\def\hhhS#1{\underline{\hat{S}_{#1}}}

\def\S(#1){{{S}_{#1}}}
\def\Ss(#1,#2){{{S}_{#1,#2}}}
\def\Sss(#1,#2,#3){{{S}_{#1,#2,#3}}}
\def\Ssss(#1,#2,#3,#4){{{S}_{#1,#2,#3,#4}}}
\def\Sssss(#1,#2,#3,#4,#5){{{S}_{#1,#2,#3,#4,#5}}}

\def\cP{{\cal{P}}}
\def\gone{\cP}
\def\dgone{\dot{\cP}}
\def\ddgone{\ddot{\cP}}
\def\dddgone{\dddot{\cP}}

\def\np#1#2#3{{Nucl.\ Phys.\ }{\bf B#1} (#3) #2}
\def\pl#1#2#3{{Phys.\ Lett.\ }{\bf B#1} (#3) #2}

\title{
{${\cal{N}}\!=\!4$ SUSY Yang--Mills: three loops made simple(r) }} 
\author{Yu.~L.\ Dokshitzer$^1$\footnote{On leave of absence: St.\
    Petersburg Nuclear Physics Institute, 188350, Gatchina, Russia} and
  G.~Marchesini$^{2}$ 
  \\
  \normalsize
  $^1$LPTHE, Universities of Paris-VI and VII and CNRS, Paris, France\\
  \normalsize $^2$University of Milano--Bicocca and INFN Sezione di
  Milano--Bicocca, Milan, Italy}

\begin{document}

\maketitle

\vspace{-8.3cm}
\begin{flushright}
  Bicocca--FT--06--20\\
  hep-th/0612248
\end{flushright}
\vspace{5.5cm}

\abstract{We construct universal parton evolution equation that 
produces space- and time-like anomalous dimensions for
the maximally super-symmetric ${\cal{N}}\!=\!4$ Yang--Mills field theory model, 
and find that its kernel 
satisfies the Gribov--Lipatov  reciprocity relation in three loops.
Given a simple structure of the evolution kernel, this should help to 
generate the major part of multi-loop contributions to QCD anomalous dimensions,
due to classical soft gluon radiation effects. 
}

%

\section{Introduction}

Super-symmetric Yang--Mills theories, 
which neighbour QCD in their physical content but are easier to handle,
may play an important r\^ole in shedding light upon multi-loop QCD results. 
In this work we employ the ${\cal{N}}=4$ super-symmetric Yang--Mills (${{\cal N}}\!=\!4$ SYM) 
field theory in order to 
better understand some important features  of the QCD parton evolution.

\paragraph{Soft gluons and the ``inheritance'' idea.}
Permanent fight for increasing precision of QCD predictions necessitates higher 
order perturbative calculations. 
Such calculations, and markedly the recently completed three loop analysis 
of DIS anomalous dimensions \cite{MVV}, 
yield lengthy expressions reach in hidden physical content that one has to grasp. 
A higher loop expression necessarily contains certain structures {\em inherited}\/ from 
the preceding  orders of the PT expansion. This is the case, for instance, for the 
renormalization effects that build up the physical coupling $\alphaphys$ 
determining in all orders the intensity 
of soft gluon emission given by the Born level expression \cite{CMW,DKT}. 

We expect this ``inheritance'' feature to  
go deeper and to include, in particular, 
real and virtual radiative effects due to multiple soft gluons. The basis for such expectation is the 
celebrated, virtually unknown, Low--Burnett--Kroll theorem \cite{LBK} which 
fully applies to QCD as well as to SYM quantum field theories. 
It tell us that  
soft gluon radiation has in fact {\em classical}\/ nature.
It is independent of the {\em quantum}\/ state of the radiating parton system, 
and does not change it. 
Therefore it should be possible to find a procedure for iteratively  ``dressing''  
an underlying lower order hard parton splitting process by soft gluons. 
The realisation of this programme  would allow one to generate, by simple means, the 
 corrections that the anomalous dimension receives from soft gluons in {\em all orders}\/  
 of the perturbative expansion. 
There is a good reason to believe that soft gluon radiation effects are responsible 
for the major part of high order contributions to the QCD anomalous dimension.  
At the two loop level this has been explicitly demonstrated in a particular example of 
the time-like heavy quark fragmentation function in \cite{DKT}. 

To verify this expectation at the three loop level we use as a testing ground the ${{\cal N}}\!=\!4$ SYM model.
Since soft radiation is universal, the results of the present study are directly applicable to QCD 
(modulo adjustment of colour factors).
Indeed, the diagonal one loop anomalous dimensions ($q\to q$, $g\to g$) 
contain the universal {\em classical}\/ part of the gluon emission probability, 
\beq\label{eq:LBK}
 \frac{dw}{dx} = C_p \frac{\alphaphys}{\pi} \left[ \frac{x}{1-x} + (1-x)g_p(x)\right] .
\eeq
Being proportional to the ``colour charge squared'' $C_p$, 
it is otherwise insensitive to the nature of the radiating particle $p$.  
Quantum effects encoded in $g_p$ obviously depend on $p$
($g_{1/2}=\half$ for a fermion, $g_1=x+x^{-1}$ for a vector emitter, etc.). 
Importantly, as a consequence of the LBK theorem,
the  genuine {\em quantum}\/ contribution to \eqref{eq:LBK} 
is down, relative to the universal classical piece,  
by {\em two powers}\/ in the $(1\!-\!x)$ counting. 
In the standard approach, in higher orders classical and quantum effects get mixed by the $Q^2$ evolution, resulting in cumbersome expressions for  anomalous dimensions.  
  
In the ${{\cal N}}\!=\!4$ SYM theory, $g_p=0$ in \eqref{eq:LBK} at one loop: 
quantum effects due to boson and fermion fields 
compensate each other, leaving us with a pure ``classical 
physics'' at the Born level, and thus provide a perfect ground for testing the ``inheritance'' idea. 
The very idea of {\em extraction}\/ of the third loop contribution 
to the ${{\cal N}}\!=\!4$ SYM anomalous dimension \cite{KLOV} from the corresponding QCD result \cite{MVV} 
was based on the observation that the two-loop expression for the anomalous dimension 
$\gamma_{\mbox{\scriptsize SYM}}$ 
is built of the functions characterised by the same level of {\em transcedentality}\/ \cite{KL2loop}. 
In fact, such functions (with transcedentality $\tau=2k-1$ in the k$^{\mbox{\scriptsize th}}$ order of 
perturbation theory) are nothing but finite ``leftovers'' of the real--virtual cancellation of infrared 
gluon divergences.    Bearing this in mind, the study of higher orders in
the ``universal anomalous dimension'' of the ${{\cal N}}\!=\!4$ super-multiplet of twist-two 
operators  \cite{KLOV} 
should help us to understand the structure of the iteration of pure classical gluon emissions in QCD.

\paragraph{Analytic link between parton distribution and fragmentation functions.}
It is important to stress that apart from being large, 
soft gluon effects are also responsible for generating enhanced  $\ln^n(1-x)$ corrections 
in the quasi-elastic kinematics, $1-x \ll  1$, 
both for parton distribution functions in space-like process, $x_B\to 1$, 
and fragmentation functions in time-like parton cascades,  $x_F\to 1$.
Such corrections translate into $\ln N$ terms in the asymptotics of anomalous dimensions at large $N$. 
In one loop \cite{GL,AP,YD}, 
where $\gamma(N) \simeq C_p \frac{\alpha_s}{\pi} \psi(N) \propto \ln N$, 
the task of analytic continuation was solved  in \cite{BLP} where it was proposed
to treat certain logarithms 
in {\em arithmetic sense}, $\ln(1-x)\Rightarrow \ln\abs{1-x}$; in the
QCD context this idea had been implemented in \cite{YD}.
However, in higher orders the power $n$ of the $(\ln N)^n$ enhancement 
in the anomalous dimension increases, thus
impairing the analytic link between DIS and $\ee$ annihilation channels. 
For a thorough discussion of the problem at the two--loop level see \cite{VS} and
references therein.

The ${{\cal N}}\!=\!4$ SYM framework is perfectly suited for shedding light on
the problem of finding a way to analytically relate these  cross-channel processes. 
For the benefit of the reader, let us recall this problem in more detail. 

In the DIS case the large virtual momentum $q$ transferred from the
incident lepton to the target nucleon with momentum $P$ is space-like,
$q^2<0$. Inelasticity of the process is conveniently characterised by the
Bjorken variable $x_B= -q^2/2(Pq)$.
Inclusive fragmentation of an $\ee$ pair with total momentum $q$
(large positive invariant mass squared $q^2$) into a final state hadron 
with momentum $P$ is characterised by the
Feynman variable $x_F=2(Pq)/q^2$ (hadron energy fraction in the $\ee$
cms.).  The fact that Bjorken and Feynman variables are indicated by
the same letter is certainly not accidental.  In both channels $0\le x
\le 1$ though these variables are actually mutually reciprocal, $x_F
\Longleftrightarrow 1/x_B$, rather than identical.  One $x$ becomes
the inverse of the other after the crossing operation $P_\mu \to
-P_\mu$: \beq\label{eq:xes} x_B \>=\> \frac{-q^2}{2(Pq)}, \qquad x_F
\>=\> \frac{2(Pq)}{q^2}.  \eeq Apart from the difference in the hadron
momentum $P$ belonging to the initial state in the DIS and final state
in the $\ee$ case, the Feynman diagrams for the two processes are
basically the same.  ``Mass singularities'' that emerge when
additional momenta in the Feynman diagrams become collinear to $P$ are
therefore also the same. That is why in the two processes a similar
parton interpretation emerges in terms of QCD evolution equations.

The common structure of Feynman diagrams and the reciprocity
\eqref{eq:xes} are the origin of the close relation between deep
inelastic structure functions and $\ee$ annihilation inclusive
fragmentation functions.  In particular, the space- and time-like
evolution anomalous dimensions turn out to be related.

\paragraph{Drell-Levy-Yan relation. }
Back in 1969, Drell, Levy and Yan have proposed that the two channels can be linked by 
analytical continuation \cite{DLY}. 
If one takes the space-like splitting function $\tilde{\gamma}^{(\mbox{\scriptsize S})}(x)$ 
--- the Mellin transform of the anomalous dimension $\gamma^{(\mbox{\scriptsize S})}(N)$ 
--- and {\em analytically continues}\/ it from the physical region
$x\equiv x_B\le 1$ to $x>1$, such a procedure would yield the time-like splitting function
 $\tilde{\gamma}^{(\mbox{\scriptsize T})}(x_F)$ with $x_F=x^{-1}< 1$:
 \beq\label{DLY}
  \tilde{\gamma}^{(\mbox{\scriptsize T})}(x) \>=\> - x^{-1}\> 
  \tilde{\gamma}^{(\mbox{\scriptsize S})} \left(x^{-1}\right) .
 \eeq

\paragraph{Gribov-Lipatov reciprocity.}
Gribov and Lipatov in their first systematic study of DIS and $\ee$ annihilation processes in the
QFT framework \cite{GL}, 
suggested that the common dynamics in the two channels results simply in the identity  
\bsub\label{GLrecip}
\beq\label{GL}
  \tilde{\gamma}^{(\mbox{\scriptsize T})}(x) \>=\>  \tilde{\gamma}^{(\mbox{\scriptsize S})}(x) .  
\eeq
Taken together with the DLY relation \eqref{DLY} this leads to the relation known as the 
GL {\em reciprocity}\/:
\beq
  \tilde{\gamma} (x) \>=\> - x\> \tilde{\gamma}  \left(x^{-1}\right) .
\eeq
\esub
In the Mellin moment space, 
\beq
   \gamma(N) = \M{\tilde{\gamma}(x)}  \equiv \int_0^1 \frac{dx}{x}\, x^N \> \tilde{\gamma}(x), 
\eeq
\eqref{GLrecip} 
translates into the symmetry $N\to -(N+1)$; in other words,
in order to satisfy the GL reciprocity, the anomalous dimension $\gamma(N)$  
has to be a function of a single variable $N(N+1)$ --- the Casimir operator
of the collinear $SO(1,2)$ subgroup of the 
$SO(2,4)$ conformal group \cite{BKM}.
A group symmetry nature of the GL reciprocity was advocated 
already in the early 70's when the issue was first raised \cite{FGP}.
 Indeed, since $x$ is a light-cone momentum fraction,
$y= \ln x$ represents parton (pseudo)rapidity: $k_0= m_\perp \cosh y$, $k_z=m_\perp \sinh y$. 
Therefore in the Mellin transformation,
$$
  \int \frac{dx}{x}\, x^N \>=\> \int dy\, e^{yN}  \> \Longrightarrow \> \int d\phi\, e^{i\phi N}, \quad \phi= i y, 
$$
an analogy between the Lorenz boost parameter $y$ and an imaginary rotation angle 
allows one to look upon the conjugate variable $N$ as ``angular momentum'', with
${\bf J}^2=N(N+1)$ the corresponding Casimir operator. 
 

True in one loop (leading logarithmic approximation), 
see e.g.\ \eqref{eq:LBK}, the relations \eqref{GLrecip} 
are known to break already in the second loop \cite{CFP,FP80}.

\paragraph{Reciprocity respecting evolution equation.}
This letter presents a development of the project that aims to
reorganise the QFT parton evolution picture \cite{Lipatov1974} in such a way as to preserve
a close relation between space- and time-like parton dynamics.
In \cite{DKT,DMS}  a ``reciprocity respecting'' equation (RRE)
for QCD anomalous dimensions has been suggested, based on a notion of the
universal ``evolution kernel'' $P(x,\as)$, identical for two channels, 
that satisfies the Gribov--Lipatov relation in all orders:
\beq\label{eq:RRE0}
\partial_{\tau} D^{(\mbox{\scriptsize T/S})}(N,Q^2) \equiv \gamma_\sigma(N,\as)
\,D^{(\mbox{\scriptsize T/S})}(N,Q^2) \> =
\int_0^1\frac{dz}{z} \, z^N\> P(z, \as) \> D^{(\mbox{\scriptsize T/S})} \left(N,z^\sigma  Q^2\right) .
\eeq
Here $\tau=\ln Q^2$, while $\sigma=+1$ for $e^+e^-$ annihilation parton fragmentation functions 
$D^{(\mbox{\scriptsize T})}$ (time-like evolution), 
and $\sigma=-1$ for DIS parton distributions $D^{(\mbox{\scriptsize S})}$ (space-like).
Using the Taylor expansion trick we obtain the formal solution of
\eqref{eq:RRE0}:
\beq\label{eq:shift} 
\partial_{\tau}
 D^{(\mbox{\scriptsize T/S})}(N,Q^2) = \int_0^1\frac{dz}{z} z^N P(z, \as)
z^{\sigma\partial_\tau} D^{(\mbox{\scriptsize T/S})}(N,Q^2) .
\eeq
At the three-loop level, this equation has been used to analyse the  $x\to 1$ behaviour 
of the splitting functions $\tilde{\gamma}^{(\mbox{\scriptsize T/S})}(x)$ in \cite{DMS}. 
There the all-order relation $C=-\sigma A^2$
between the magnitudes of the $A(1-x)^{-1}$ and $C\ln(1-x)$ singularities was established,
observed by Moch, Vermaseren and Vogt  
for quark and gluon space-like anomalous dimensions \cite{MVV}.
The prediction of \eqref{eq:shift} for the time-like case was verified at the third loop by 
explicit calculation of the non-singlet anomalous dimension in \cite{MMVconfirm}.

The analysis of next subleading contributions in the $x\to1$ limit attempted in \cite{DMS} 
was complicated by the presence of scheme dependent  contributions 
proportional to the $\beta$ function\footnote{We thank G.\ Korchemsky 
for communicating us an elegant solution of the problem encountered in \cite{DMS}.}, 
starting from the constant term 
$\tilde{\gamma}(x)=\cO{1}$, which contribution also had to be under full ``inheritance''
control, according to the LBK wisdom. It is indeed 
in the ${{\cal N}}\!=\!4$ SYM model, having $\beta(\as)\equiv 0$.  

Here the RRE \eqref{eq:shift} 
readily reduces to a pair of complementary relations which hold  {\em in all orders}\/ and for arbitrary $N$:
\bsub\label{RRE}
\beeq\label{eq:2chan}
\gamma^{(\mbox{\scriptsize T/S})}(N)  &=& \gone\big(N 
+\sigma\cdot \gamma^{(\mbox{\scriptsize T/S})}(N) \big), \\
\gone(N)  &=& \gamma^{(\mbox{\scriptsize T/S})}\big(N -  \sigma\cdot \gone(N) \big). 
\eeeq
\esub
Expanding \eqref{eq:2chan} results in
\begin{eqnarray}
  \label{eq:new-split-mellin-expand}
\gamma \equiv   \gamma^{(\mbox{\scriptsize T/S})}[\al] 
&=&    \gone + \sigma\gamma  \, {\dgone}  + \half \gamma^{2} \,{\ddgone }  
 + \sigma \textstyle{\frac1{3!}}\gamma^{3}  \,{\dddgone } + \cO{\as^5},
\end{eqnarray}
where dots indicate derivatives with respect to $N$ and $\gone \!\equiv\! \gone(N)$.  
Since $\gamma = \cO{\as}$, this equation gives iteratively a tower of inherited higher 
order contributions to the anomalous dimensions in both channels. 
Solving \eqref{eq:new-split-mellin-expand} perturbatively gives
\bsub\label{eq:RVeq}
\beeq
\label{eq:Req}
 \half(\gamma^{(\mbox{\scriptsize T})} + \gamma^{(\mbox{\scriptsize S})}) 
 & \equiv&  \gone+ R, \qquad R=   \left[  \gone  \dgone^{2} 
+ \half \gone^{2} \ddgone    \right] + \cO{\as^5} , \\
\label{eq:Veq}
 \half(\gamma^{(\mbox{\scriptsize T})} - \gamma^{(\mbox{\scriptsize S})}) 
 &\equiv&    V   = \gone\dgone + \bigl[ \gone \dgone^3 + \threehalf 
 \gone^2\dgone\ddgone + \textstyle{\frac16}\gone^3\dddgone \bigr]+ \cO{\as^6}. 
\eeeq
\esub
The difference of space- and time-like anomalous dimensions $V$
in \eqref{eq:Veq} does not have a linear term in the evolution kernel $\gone$; therefore,   
knowing $\gone$ at order $k$, one predicts $V$ at the order $k+1$.  
Moreover, since the induced terms in \eqref{eq:RVeq} contain powers of $\gone$,  
the lowest order evolution kernel $\gone_1 \propto \ln N$ generates a tree of logarithmically enhanced terms, up to $\as^k \ln^k N/N^{k+1}$ in order $k$ of the perturbative expansion.

If the large-$N$ singularity of the evolution kernel $\gone$ turns out to be stable under perturbation,
given $\gone$ at order $k$, \eqref{eq:new-split-mellin-expand} would predict all 
logarithmically enhanced contributions $f_m(N) \cdot \ln^m N $, $2\le m \le k+1$ 
at the next order $k+1$, where  $f_m(N) =\cO{N^{-k-1}}$ are known functions that 
are regular at $N=\infty$.

\medskip

We translate the space-like ${{\cal N}}\!=\!4$ SYM anomalous dimension recently calculated by 
Kotikov, Lipatov, Onishchenko and Velizhanin  in three loops \cite{KLOV} into the 
corresponding evolution kernel $\gone$ of  \eqref{RRE}. 
We find that $\gone_3$ is given by a rather transparent expression, 
much more compact than $\gamma_3$, 
exhibits only a single power of the $\ln N$ enhancement at large 
$N$ and satisfies the GL reciprocity \eqref{GLrecip}.
Grasping internal logic of the second and third order contributions to $\gone$ 
should help to gain better understanding of the structure of multi-loop QCD anomalous dimensions.

\bigskip
The paper is organised as follows. In section \ref{Sec-Answer} we present 
the evolution kernel $\gone$ in three loops. In section \ref{Sec-Disc}  we expose in detail 
essential properties of the answer, including important kinematical limits. 
In section \ref{Sec-Conc} we present our conclusions and discuss perspectives for future studies.

\section{Answer \label{Sec-Answer}}

We construct the RR evolution kernel,
\beq\label{series}
 \gone(\al,N) = \sum_{k=1} a^k\, \gone_k(N), \quad
  \gamma^{(\mbox{\scriptsize T/S})}(\al,N) = \sum_{k=1} a^k\, \gamma_k^{(\mbox{\scriptsize T/S})}(N),
 \qquad 
  a=\frac{C_A\al}{\pi}, 
\eeq
employing the space-like anomalous dimension, $\gamma=\gamma^{(\mbox{\scriptsize S})}$, 
for the ${{\cal N}}\!=\!4$ SYM model. 
From the perturbative expansion of \eqref{eq:RVeq}, 
\beq
\begin{split}
\gamma_1 &= \gone_1 , \quad
\gamma_2  \>=\> \gone_2 - \gone_1\dgone_1 ,\\
\gamma_3 &= \gone_3 -
 \left [\dgone_1\gone_2 + \gone_1\dgone_2 \right] 
 +  \left[  \gone_1 \dgone_1^{2} + \half \gone_1^{2} \ddgone_1  \right] .
\end{split}
\eeq
Within our normalisation convention \eqref{series}, 
one has $\gamma_k = \guni^{(k-1)}/4^k$, with $\guni$ the anomalous dimension 
in the original notation of  \cite{KLOV}. 
The latter is given in terms of {\em harmonic sum}\/ functions  $S_{\pm a}(N)$ and $S_{-p,a, \ldots }(N)$,
see \cite{RemVer,Blum00} and Appendix \ref{Aharm}. 
It reads
\bsub\label{eq:uni}
\beeq
\gamma_1 &=& - S_{1} ;\\
\label{eq:uni1}
\gamma_2 &=& 
 \half S_3 +  S_1S_2 + \left( \half  {S}_{-3} + S_1 {S}_{-2} -  {S}_{-2,1} \right) ; \\
\label{eq:uni2}
\gamma_3 &=&
 -\half S_5  - \left[ S_1^2S_3 +\half S_2S_3 +S_1S_2^2 
+\threehalf S_1S_4 \right] \nonumber  \\[1mm]
\label{eq:Zdef}
 && - S_1\left[\, 4 {S}_{-4} +\half  {S}_{-2}^2 +2S_2 {S}_{-2}
  -6 {S}_{-3,1} -5 {S}_{-2,2} +8  {S}_{-2,1,1}\, \right] \\
&& - ( \half S_2 + 3S_1^2)  {S}_{-3} -S_3 {S}_{-2} + (S_2+ 2S_1^2)  {S}_{-2,1} 
 +12 {S}_{-2,1,1,1} \nonumber  \\
 \label{eq:Ydef}
&&   -6( {S}_{-3,1,1}+ {S}_{-2,1,2}+ {S}_{-2,2,1}) 
+ 3( {S}_{-4,1}+ {S}_{-3,2}+ {S}_{-2,3})  - \threehalf  {S}_{-5} . \nonumber
\eeeq
\esub
In order to arrive at a compact expression for the third loop evolution kernel $\gone_3$
we introduce the characteristic function $\varphi$, 
\bsub\label{YmNdef}
\beq\label{phidef}
 d\varphi = 
 \frac{dz}{z}\,\frac{z-1}{z+1}, \qquad  \varphi(z) =  \ln \frac{(1+z)^2}{4\,z} = \ln \cosh^2(\half\ln z),
\eeq 
which is invariant under the transformation $z\to z^{-1}$,  
construct integrals 
\beq\label{Phidef}
   \Phi_\tau (x) \>=\> \frac1{\Gamma(\tau)} \int_x^1 \frac{dz}{z}\, 
   \big(\varphi(x)- \varphi(z)\big)^{\tau-1} ,
\eeq
(where subscript $\tau$ marks transcedentality of the function),  
and define Mellin transforms 
\beq\label{Ydef}
  \hat{Y}_{-m}(N) \>=\> {(-1)^N} \, \M{\frac{ x}{1+x} \, \Phi_{m-1}(x)  } .
\eeq
Functions $\hat{Y}_{-m}(N)$ are given by linear combinations  
\beq\label{Ys}
\begin{split}
\hat{Y}_{-3}  &=   \hS{-3}- 2\hhhS{1,-2}  ; \qquad 
\hat{Y}_{-4}  =   \hS{-4} - 2(\hhhS{1,-3} + \hhhS{2,-2}) + 4\hhhS{1,1,-2}  ; \\
 \hat{Y}_{-5} &=   \hS{-5}  - \!  2\big( \hhhS{3,-2}\!  +\!  \hhhS{2,-3} + \hhhS{1,-4} \big)  
+ \!   4 \big(\hhhS{2,1,-2} \! +\!  \hhhS{1,2,-2}\!  +\!  \hhhS{1,1,-3}  \big)  - 8 \hhhS{1,1,1,-2}  , \quad { }
\end{split}
\eeq
\esub
where {\em hat}\/ crowns a sum with subtracted $N=\infty$ value:
\beq\label{hatdef}
 \hS{\vec{a}}(N)\>\equiv S_{\vec{a}}(N) -  S_{\vec{a}}(\infty).  
\eeq
Here $\hhhS{\vec{m},-p}(N)$ are {\em complementary harmonic sums},  
\beq\label{Scompl}
 \hhhS{a_1, a_2,\ldots , a_n}(N) = (-1)^n \!\! 
 \sum_{k_1=N+1}^\infty \!\!\frac{(\sgn{a_1} )^{k_1}}{k_1^{\abs{a_1}}} 
 \sum_{k_2=k_1+1}^\infty \!\!\frac{(\sgn{a_2} )^{k_1}}{k_1^{\abs{a_2}}}\>\>
 \cdots \!\!\! \sum_{k_n=k_{n-1}+1}^\infty \!\!\frac{(\sgn{a_{n}} )^{k_n}}{k_1^{\abs{a_n}}} ,
\eeq
whose relation to standard multi-index harmonic sums 
is displayed in Appendix \ref{Aharm}. 
The functions \eqref{Ys} are built of harmonic sums with  one, and the rightmost, 
 {\em negative index}\/ (therefore the notation $Y_{-m}$). 
Internal logic 
of the constructs \eqref{Ys} is transparent and makes generalisation straightforward.

In these terms, the ${{\cal N}}\!=\!4$ SYM evolution kernel, 
 in first three orders of the perturbative expansion in the physical coupling, 
\beq
\alphys =  a \left(1 -\half\zeta_2 a +{\textstyle \frac{11}{20}}\zeta_2^2 a^2 + \ldots \right), 
\eeq
takes a compact form
\bsub\label{P123}
\beeq\label{P1}
 \gone_1 &=& - \> S_1; \\ 
\label{P2}
  \gone_2 &= &\>\>\> \half  \hS{3}  - \half \hat{Y}_{-3}       \> + B_2; \\
\label{P3}
\gone_3 &= &  -\half  \hS{5} + \threehalf \hat{Y}_{-5} 
   \, + B_3 \> 
+\,   \zeta_2 \cdot \half \hS{3}  \nonumber\\
&& + \>S_1 \cdot \left[ \hat{Y}_{-4} \>- \half\big( \hS{-4} +  \hS{-2}^2 \big) 
\>\>+\> \zeta_2\cdot \half \hS{-2}   \right]  . 
\eeeq
\esub
Here
$ B_2= \textstyle{\frac34}\zeta_3$, 
$ \> B_3=  - {\textstyle\frac18}\zeta_2\zeta_3  -{\textstyle \frac54} \zeta_5$.

\section{Discussion  \label{Sec-Disc}}

Let us list some characteristic features of the RR evolution kernel  \eqref{P123}.

The evolution kernel \eqref{P123} (as well as anomalous dimensions) 
contains harmonic sums with positive and negative indices. These contributions have different 
origin and play markedly different r\^oles in extreme kinematical limits.  
Harmonic sums bearing a negative index (indices) 
oscillate with $N$, see  \eqref{Sneg}  and \eqref{Ydef}: 
\beq\label{posneg}
  \gone(N) \>=\>  {\gone}^{\mbox{\scriptsize pos}} (N) \,+\, (-1)^N\cdot 
  {\gone}^{\mbox{\scriptsize neg}} (N) .
\eeq
The presence of the oscillating factor $(-1)^N$ calls for introducing two separate functions, 
analytically continued from even and odd moments $N$. Indeed, in QCD one has two independent 
non-singlet anomalous dimensions, replacing \eqref{posneg} by 
\beq
 \gamma^{\,(\pm)}_{\,\rm ns} \> = \> p_{qq}\> \pm\>  p_{q\bar{q}} .
\eeq 
The two anomalous dimensions have different {\em signature}\/ with respect 
to $s\leftrightarrow u$ ($x\to -x$) crossing. In ${{\cal N}}\!=\!4$ SYM ``quarks'' are Majorana 
fermions, making $q$ and $\bar{q}$ physically indistinguishable. 
In the light-cone formulation of ${{\cal N}}\!=\!4$ SYM \cite{Mand,BLN},
in the study of anomalous dimensions \cite{BDKM}
the negative signature does not appear \cite{BDKM-2},
since operators built of the unique superfield of the theory
have vanishing {\em odd}\/ moments.
Therefore, in the ${{\cal N}}\!=\!4$ SYM context the only possible continuation to non-integer moments 
consists of dropping the factor $(-1)^N$, as it has been done in \cite{KLOV},
resulting in $\guni \equiv \gamma^{(+)}$. 
We should keep in mind, however, that both signatures are present in the QCD case, where 
the maximal transcedentality structures of \eqref{P123} are also present and are responsible for 
a significant part of higher order contributions.

\subsection{Inheritance}
Compared to the standard anomalous dimension \eqref{eq:uni}, the presence of harmonic sums with {\em positive indices}\/ in the RR evolution kernel \eqref{P123} is minimal: 
\beq\begin{split} \label{inher}
\gamma_2 &=  \half S_3 +  S_1S_2 \,+ \cdots , 
\qquad\qquad \qquad\qquad \qquad\qquad \qquad 
\gone_2 =  \half S_{3} \, + \cdots  ; \\
\gamma_3 &=
 -\half S_5  - \left[ S_1^2S_3 +\half S_2S_3 +S_1S_2^2 +\threehalf S_1S_4 \right]  \,+ \cdots ,\quad
\gone_3 =  -\half S_{5}  \, + \cdots ,
\end{split}
\eeq
where $\cdots$ mark contributions of negative index sums.
Thus, {\em all}\/ non-linear combinations of $S_i$ in the anomalous dimension has been generated 
by the RRE and are therefore physically  ``inherited'' from the {\em first loop}.

\subsection{Negative index sums: relation to Feynman graph discontinuities}

Sums with one negative index contribute to the anomalous dimension starting from the second loop.
Their appearance can be traced to {\em non-planar}\/ diagrams that can be cut 
both in $s$ and $u$, and thus have non-zero {\em Mandelstam double spectral function},
$\rho_{su}\neq 0$.

\begin{figure}[h]
\begin{center}
\epsfig{file=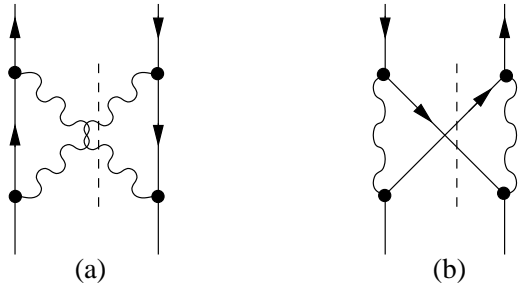}
\end{center}
\caption{On the origin of negative index harmonic sums\label{cross-fig}}
\end{figure}

The diagrams for non-singlet QCD quark evolution shown in Fig.~\ref{cross-fig} can be related by the 
crossing transformation $s \leftrightarrow u\simeq -s $  which, in the language of the Bjorken $x$ 
variable, formally corresponds to the reflection $x \to -x$.
It is therefore only natural that the {\em second loop}\/ 
non-singlet $q_i\to \bar{q}_i$ anomalous dimension described by the graph (b),  is proportional, 
in $x$ representation, to the {\em first loop}\/ $q_i\to q_i$ splitting function 
evaluated at the reflected point, 
\beq
{p} 
_{q\bar{q}}(x) \>=\> \alphys^2\, (\half C_A-C_F)\> p
_{qq}(-x)\cdot \Phi_2(x) , \quad  
p
_{qq} (x)=\frac{1+x^2}{2(1-x)}.
\eeq
The proportionality factor $\Phi_2(x)$ 
is a function\footnote{called $S_2(x)$ in the original papers 
\cite{CFP},  \cite{FP80}}  of transcedentality 2, defined in \eqref{Phidef}. 
It is important to stress that this factor appears, invariably, 
in second loop anomalous dimensions, multiplying the full first loop splitting functions
$p_{BA}(-x)$ in {\em all  parton transitions}, $B\to A$,  including non-diagonal ones, 
see \cite{FP80}.
Using three--loop results of \cite{MVV}, it is straightforward to verify  that in the non-singlet QCD
anomalous dimensions $\gamma_{\rm ns}^{(\pm)}$ {\em every}\/ harmonic sum 
(maximal transcedentality or not)  bearing a negative index is proportional to $2C_F\!-\!C_A$  
(colour suppressed), that is has  non-planar origin.

Let us compare contributions of positive and negative index sums.
For the second loop evolution kernel we have
\bsub
\beeq
\Min{ \>\half \hS{3}\>} &=& \>\>\> \half\> \tilde{\gone}_1(x)\>\>\cdot \> \big( -\half \ln^2x \big)  
\equiv    \tilde{\gone}^{\mbox{\scriptsize pos}} _2(x) , \\
 \Min{-\half \hat{Y}_{-3}\cdot (-1)^N} &=&  
- \half  \, \tilde{\gone}_1(-x) \cdot \big( -\Phi_2(x)\, \big)
\equiv \>   \tilde{\gone}^{\mbox{\scriptsize neg}} _2(x) , 
\eeeq
\esub
with $\tilde{\gone}_1(x)=x/(1-x)$ the one loop splitting function, 
the inverse Mellin image of $\gone_1(N) = -S_1(N)$. 
The fact that these two terms originate from different cuts of the same diagram can be seen 
 from  internal structure of corresponding contributions:
\bsub\beeq
  \tilde{\gone}^{\mbox{\scriptsize pos}} _2(x)  
&=& \half \,\frac{x}{{\bf [1 - x]}} \cdot  \int_x^1 \frac{dz}{z}\, \frac{{\bf [1 + z]}}{1+z} \ln z , \\
  \tilde{\gone}^{\mbox{\scriptsize neg}} _2(x)  
&=& \half   \,\frac{x}{{\bf [1 + x]}}  \cdot \int_x^1 \frac{dz}{z}\, \frac{{\bf [1-z]}}{1+z} \ln z.
\eeeq\esub
where we have used \eqref{phidef} to transform the integral
$$
  \Phi_2(x) \equiv  \int_x^1 \frac{dz}{z}\,  \ln\left(\frac{(1+x)^2\,z}{(1+z)^2\,x}\right) 
  = \int_x^1 \frac{dz}{2z} \,\frac{z-1}{z+1}\,  \ln z.
$$
Higher order {\em positive index}\/ contributions, 
$\hS{2k-1}$ in \eqref{P123}a,b, with $k$ the order of the perturbative
expansion, have a simple origin and a simple structure:
\bsub\beeq
 \hS3(N) &=& \frac1{2!}\, \frac{d^2}{dN^2} S_1(N), \qquad \Min{\hS3} = \frac1{2!}\, \frac{x\, \ln^2x}{x-1};\\
 \hS5(N) &=& \frac1{4!} \, \frac{d^4}{dN^4} S_1(N), \qquad \Min{\hS5} = \frac1{4!}\, \frac{x\, \ln^4x}{x-1}.
\eeeq\esub
Given that the negative and positive index contributions are related by 
the $s\leftrightarrow u$ crossing, and guided by the second loop example, 
a possibility to {\em generate}\/ the negative index piece from its positive counterpart  
does not look implausible.

\subsection{Gribov--Lipatov reciprocity}

The Gribov--Lipatov reciprocity is respected by each piece entering the evolution kernel.
Indeed, it is easy to see from  \eqref{SMel}  that the inverse Mellin images of
harmonic sums $\hS{a}$ with {\em odd}\/ $a$, 
as well as negative index sums $\hS{-b}$ with $b$ {\em even}, satisfy the relation
\beq\label{x1x}
    F(x)\>=\> -  x\, F(x^{-1})
\eeq
(For the distribution $\Min{S_1}$ this relation holds for $x\neq1$.) 
The same is true for functions $\hat{Y}_{-m}$, see \eqref{YmNdef},
\bsub\beq
\Min{(-1)^N\hat{Y}_{-m}}  = \frac{x}{1+x}\cdot \Phi_{m-1}(x), 
\eeq
which transform according to \eqref{x1x},  due to 
\beq
     \Phi_\tau(x^{-1}) \>=\> - \,\Phi_\tau(x).
\eeq
\esub
Finally, an observation that \eqref{x1x} is stable under {\em convolution}\/ 
of two RR functions, concludes the proof that every item in \eqref{P123} satisfies 
the GL reciprocity relation   \eqref{GLrecip}.

In the moment space, as we have discussed above, one expects symmetry with respect 
to the substitution 
\beq\label{symm}
\gone(N) \>=\> \gone( -N-1), 
\eeq
and thus the dependence on the combination $N(N+1)$.
Indeed, in QCD, for example, the second loop non-singlet positive signature evolution 
kernel reads \cite{DKT}   
\beq
\begin{split}
{\tilde{\gone}^{({\rm ns},+)}_{2}(x)} &= (\half C_A-C_F)\big[\,  p
_{qq} (x)\cdot \half \ln^2x 
 + p
 _{qq} (-x)\cdot \Phi_2(x)\, \big]  \\
 & -\quart C_F\big(\half(1\!+\!x)\ln x + (1\!-\!x) \big) + \quart C_A(1\!-\!x). 
 \end{split}
\eeq
It contains ``algebraic'' pieces whose dependence on the ``Casimir'' is transparent,
\beeq
 \M{1-x} = \frac1{N(N+1)}, \quad \M{(1+x)\ln x} = \frac{2N(N+1)+1}{N^2(N+1)^2}.
\eeeq
For {\em maximal transcedentality}\/ contributions this property also holds, in certain  
sense. It cannot be applied to integer $N$ since the reflected image of, say, 
the leading harmonic sum $S_1(N)$ has poles at each positive integer point:
\beq
S_1(-N-1)  \>=\> S_1(N) \>-\> \frac{\pi\cos\pi N}{\sin\pi N} .
\eeq
However, the symmetry  holds for {\em half-integer}\/ values of $N$, from where the
relation \eqref{symm} can be continued onto the entire complex $N$-plane.

\subsection{Large $x$}

Quasi-elastic scattering, $(1-x)\ll 1$, is determined by the behaviour at large $N$, where
\bsub
\beeq
\label{YlargeN}
\hS{a}(N) &\simeq& {N^{-a+1}}, \quad \hS{-a}(N) \>\simeq\>  \half {N^{-a}}, \\
\hat{Y}_{-a}(N) &=& \cO{N^{-2a+2}}\qquad (a\ge 2).
\eeeq
\esub
The second loop evolution kernel \eqref{P2} is ``regular'' at $N=+\infty$ 
(that is, does not contain $\ln N$ terms) and is dominated by the
positive index term,
\beq\label{lead2}
  \hS{3}(N)\simeq N^{-2}  \>\> \Longrightarrow\>\>  
  \delta \tilde{\gone}_2(x) \>\sim\>  
  (1-x) .
\eeq
It is down by two powers in
$(1-x)$ counting as compared to the Born term $\alphys(1-x)^{-1}$, 
while the negative index contribution is suppressed ever further.
Power counting looks natural: the graph Fig.\ref{cross-fig}(b) contains in the intermediate state
two {\em hard partons}, each producing the phase space suppression factor 
$d^3k/k\propto (1-x)^2$, 
resulting in 
$$
\hat{Y}_{-3}(N) \propto N^{-4}\>\>\Longrightarrow\>\>
\delta\tilde{\gone}_2(x)\>\sim\> 
(1-x)^3\>\propto\> (1-x)^{-1}\cdot [(1-x)^2]^2 .
$$  
Unlike the second loop, the third loop contribution to the kernel $\gone_3$ 
exhibits a single power of logarithmic enhancement, $S_1(N)\propto\ln N$.  
Now the positive index term is negligible, $\hS{5}\propto N^{-4}\Rightarrow \alphys^3(1-x)^3$, and it is
the last {\em singular}\/ term  in \eqref{P3}  that turns out to be leading in this limit:
\beq\label{lead3}
 \half\zeta_2 \cdot S_1(N)\hS{-2}(N) \>\simeq\> \frac{\zeta_2}{4}\, \frac{\ln N}{N^2} 
 \>\> \Longrightarrow\>\>  
 \delta \tilde{\gone}_3(x) \sim 
 (1-x)\ln(1-x) .
\eeq
It is worth noticing that the origin of the contribution \eqref{lead3} is rather specific. 
Namely, it originates from harmonic sums with {\em two negative indices}. 
In QCD, maximal transcedentality sums of this nature in the third loop are \cite{MVV}
\beeq
 \textstyle{\frac{1}{64}} \gamma^{+}_{\,\rm ns} &\Rightarrow& \quart C_F(2\cf\!-\!\ca)\, 
\big[ -2(\cf\!+\!\ca) \Ss(-3,-2) -2 (3\cf\!-\!\ca) \Ss(-2,-3) \nonumber  \\
&& + 4 \ca \Ss(-2,-2,1)
+4(3\cf\!-\!2\ca) \Ss(-2,1,-2) + 2 
(3\ca\!-\!2\cf)   \Ss(1,-2,-2) \big]. \qquad { }
\nonumber
\eeeq
Equating the colour factors, $C_F=C_A$, 
one obtains a logarithmically enhanced contribution to
the universal ${{\cal N}}\!=\!4$ SYM anomalous dimension,
\beq
\begin{split}
 \textstyle{\frac{1}{64}} \guni \>&\Rightarrow\> 
  \Ss(-3,-2) +  \Ss(-2,-3) -  \Sss(1,-2,-2) - \Sss(-2,1,-2) - \Sss(-2,-2,1)    
= - \half \,S_1\big( S_{-2}^2 + S_4\big).
\end{split}
\eeq
The term built of positive index sums, $S_1S_4$, is taken care of by the RRE, 
while the squared negative index sum, $S_{-2}^2$, propagates to the evolution kernel \eqref{P3}:
\beq
- \half  S_1S_{-2}^2 = - \half S_1\big( \hS{-2} - \half\zeta_2\big)^2 = 
-\textstyle{\frac{1}{8}}\zeta_2^2\cdot S_1 - \half S_1\hS{-2}^2 \> + \half\zeta_2 \,S_1\hS{-2}. 
\eeq
The first term on the r.h.s.\ participates in forming the $\alpha^3$ correction to the physical coupling, 
the second is $\cO{\ln N/N^4}$. The last term is a cross-product of the subtracted sum, 
$\hS{-2}\propto N^{-2}$,  and $S_{-2}(\infty)=-\half\zeta_2$, and gives rise to a bizarre correction \eqref{lead3}.

The third loop contribution \eqref{lead3} looks as a 
radiative correction to the previous order kernel \eqref{lead2}.
Taken at face value, the presence of this logarithmically enhanced contribution, 
though subleading in the overall $(1-x)$ counting,  
negates our attempt to construct a perturbative scheme such that 
relatively large higher order corrections in all orders would be generated automatically. 
The fact that this proposal stumbled at the
level of the third loop,  calls  for a deeper insight into 
the physical origin of contributions described by harmonic sums 
with two negative indices, in order to incorporate this eventuality into
the ``inheritance'' framework by further elaborating the RR equation and its evolution kernel.

\subsection{Small $x$}

Small-$x$ behaviour of the anomalous dimension, and of the evolution kernel, is determined by
the rightmost singularity in the $N$ plane. In reality, the BFKL limit in ${{\cal N}}\!=\!4$ SYM 
is described by the anomalous dimension $\guni(N-2)$ having a pole at $N\!=\!1$. 
For the sake of simplicity, we will ignore the shift of the argument and discuss the behaviour of $\gone(N)$ at $N=-1+\omega$ for $\omega\to0$. 
Participating harmonic sums have the following expansion at this point:
\bsub\label{smr}
\beeq\label{smrp}
  S_{1}(\omega) &=& -\omega^{-1} +\zeta_{2} \cdot\omega\>+\sum_{k=3} (-1)^k\zeta_k \omega^{k-1},   \\
  \hS{a}(\omega) &=&  -\omega^{-a}  -\zeta_{a} + \cO{\omega}, \quad a\ge 2; \qquad{ }\\
\label{Sm2small}
  \hS{-m}(\omega) &=& - \hS{m}(\omega) +  \cO{\omega^0}, \qquad
   \hS{-2}(\omega) = \omega^{-2}   - \half \zeta_2  
   + \cO{\omega} ; \\
\label{YSsmall}
 \hat{Y}_{-m}(\omega) 
 & = & \hS{-m}(\omega) -\zeta_2\,\hS{-m+2} + \ldots ;
  \qquad \qquad \omega=N+1\to 0, \qquad { }
\eeeq
\esub
where the factor $(-1)^N$ has been suppressed in $\hS{-m}$ and $\hat{Y}_{-m}$. 
At small $x$, positive and negative index sums are equally important. 

According to \eqref{smr},  on the second line of  \eqref{P3} the leading and first subleading singularities 
cancel,   
\beq
\begin{split}
  \hat{Y}_{-4} \>-& \half\big( \hS{-4} +  \hS{-2}^2 \big) +\half\zeta_2\hS{-2} \\
 & \simeq \>  \big({\omega^{-4}} -\zeta_2\omega^{-2} 
 \big) 
 - \half\big( \omega^{-4} +  \left[ \omega^{-2}-\half{\zeta_2} 
 \right]^2 \big)  +\half \zeta_2\omega^{-2} 
 \>=\>\cO{\omega^{-1}} ,
\end{split}
\eeq 
and the evolution kernel can be approximated as
\bsub\label{Psmall}
\beeq
  \gone_1 &=& - \>\>\> S_1 \>\qquad \> \qquad \quad\> \>\>  \simeq \>\>\> \> 
  \omega^{-1} -\zeta_2\omega \>+\ldots, \\ 
\label{P2small}
  \gone_2 &\simeq &\>\>\> \half\hS{3} -\half\hat{Y}_{-3}
 \qquad \quad \simeq - \omega^{-3} +\half \zeta_2\omega^{-1} \>+\ldots
  , \\ 
\label{P3small}
\gone_3 &\Rightarrow &  -\half \hS{5} +\threehalf\hat{Y}_{-5} +\half\hS{3}
\simeq \>2\omega^{-5} - 2 \zeta_2\omega^{-3} 
+  \ldots\, .
\eeeq
\esub
To obtain the anomalous dimension one has to add to \eqref{Psmall} the generated 
pieces \eqref{eq:RVeq}.
From  \eqref{Psmall} one easily derives perturbative expansion coefficients for the generated
reciprocity respecting (R) and violating terms (V): 
\bsub\label{generated}
\beeq\label{generated1}
 R_2 &=& 0 , \\
 V_2 &=& \gone_1\dgone_1  \>=\>  -\omega^{-3} 
 + \cO{1}; \\
\label{genR3}
R_3 &=& 
 \gone_1\dgone^2_1 + \half \gone_1^2\ddgone_1
\>  =\>\> {2}{\omega^{-5}}  -\>  {\zeta_2}\omega^{-3} \>+ \cO{1} , \\
 \label{genV3}
 V_3 &=&  \gone_1\dgone_2 \>+\> \dgone_1\gone_2 
 \>\>\>  = \> \>
 {4} \omega^{-5} -  3\zeta_2 \omega^{-3} 
  +\cO{\omega^{-2}} .
\eeeq
\esub
The inherited contributions,
\bsub\label{RpmV}
\beeq
R_2+\sigma V_2 &=& \>\> -\sigma \omega^{-3}  
+ \cO{1}, \\
R_3 +\sigma V_3 &=& 2(1+2\sigma)\omega^{-5} - {(1+3\sigma)\zeta_2} \omega^{-3} \>+\ldots\,,
\eeeq\esub
also exhibit double logarithmic series and
contain subleading singularities.
Adding the genuine third loop evolution kernel \eqref{Psmall} and
the generated pieces \eqref{RpmV}, 
for the {\em space-like}\/ channel, $\sigma=-1$, we observe cancellation of double logarithmic terms, 
together with  {\em single logarithmic}\/ singularities: 
\beeq\label{PsmallS}
\gamma^{(\mbox{\scriptsize S})}
\>\simeq\> \frac{\alphys}{\omega} \>+\> 0\cdot \frac{\alphys^2}{\omega^2} 
\>+\> 0\cdot \frac{\alphys^3}{\omega^3}\>\> +\> \zeta_2  \frac{\alphys^2}{2\, \omega}  
+  \ldots  , \qquad \omega=N+1\ll 1,
\eeeq   
the latter corresponding to two famous ``zeroes'' in the BFKL anomalous dimension \cite{BFKL}.

For the {\em time-like}\/ evolution, $\sigma=+1$, combining \eqref{Psmall} and  \eqref{RpmV}
produces 
\beeq\label{PsmallT}
  \gamma^{(\mbox{\scriptsize T})} &  \simeq&  \> \frac{\alphys}{ \omega} 
  - 2\cdot \frac{\alphys^2}{\omega^3} + 8\cdot \frac{\alphys^3}{\omega^5} 
 \>\> +\> \zeta_2 \cdot \left( \frac{\alphys^2}{2\, \omega} - 6 \frac{\alphys^3}{\omega^3} \right)+  \ldots .
\eeeq

\subsection{RRE and coherence in small-$x$ physics}

Thus, in the small-$x$ region the evolution kernel $\gone_k$ 
contains {\em double logarithmic}\/ enhancement factors,  $(\alphys/\omega^2)^{k-1}$, 
multiplying the one loop anomalous dimension, $\alphys/\omega$. 
In fact, the leading singular contributions given by the first terms in \eqref{Psmall} 
represent the start of the all-order series, 
\beq\label{PDL}
 \gone_{\mbox{\scriptsize DL}}   = \half\bigg( \sqrt{\omega^2 + 4\alphys} -\omega \bigg)
  \simeq \frac{\alphys}{\omega} -  \frac{\alphys^2}{\omega^3} + 2\frac{\alphys^3}{\omega^5} -5\frac{\alphys^4}{\omega^7} +14\frac{\alphys^5}{\omega^9} +  \ldots ,
\eeq
that are characteristic for the evolution kernel. Its origin lies simply in the choice of the logarithmic 
ordering variable that separates successive parton splittings, known as ``parton evolution time''. 
As has been shown in \cite{DMS}, in order to unify the space-
and time-like parton dynamics one has to  choose {\em parton fluctuation  lifetime},  $k_+/k^2$, 
as  a common evolution variable for both channels. It is this choice that leads to the non-local 
evolution equation \eqref{eq:RRE0} whose universal 
kernel $\gone$  no longer coincides with the anomalous dimension. 
The latter acquires an additional term, $\sigma V$,  which is opposite in sign in the two channels. 
This specific contribution, which is solely responsible for the breaking of the GL
reciprocity, is of pure kinematical origin and is easy to derive since, 
in a given order of perturbative expansion, it is inherited from lower orders.

Inherited $R$ and $V$ contributions \eqref{eq:RVeq} also posses DL enhanced terms, 
see \eqref{RpmV}. 
They combine with those of the kernel proper \eqref{PDL} and {\em cancel}\/  in the space-like case 
to produce single logarithmic BFKL series \eqref{PsmallS}. 
The cancellation of DL enhanced terms in $\gamma^{(\mbox{\scriptsize S})}$ 
holds in all orders, the reason being a general physical observation, due to V.N.\ Gribov,
that inelastic diffraction vanishes in the forward direction.

Indeed, consider a space-like scattering process in which parton $p$
emits a soft gluon, $p\to p'+k_1$ with $k_{1+}\ll p_+$, which gluon,
in turn, radiates a still softer gluon $k_2$ ($k_{2+}\ll k_{1+}$).
The logarithmic condition, according to the fluctuation lifetime ordering,  
reads
\bsub
\beq
      \frac{{\bf k}_{1t}^2}{k_{1+}}  \><\>   \frac{{\bf k}_{2t}^2}{k_{2+}}.
\eeq
This region includes, apart from ${\bf k}_{1t}^2 < {\bf k}_{2t}^2$, a ${\bf k}_{1t}$
integration over the interval
\beq\label{eq:wrong}
   {\bf k}_{2t}^2 \><\> {\bf k}_{1t}^2 \><\> \frac{k_{+1}}{k_{+2}} \cdot {\bf k}_{2t}^2,
\eeq
\esub 
which is rather large in the soft kinematics and contributes 
as $\as\ln x\cdot \as\ln^2x$ to the {\em evolution kernel}\/ $\gone_2$.  
However, in the {\em anomalous dimension}\/  
this contribution cancels when one takes into account emission of $k_2$ off the partons $p$ and $p'$.
Physically, the process can be viewed upon as break-up ({\em inelastic diffraction}\/)  
of the incident particle $p$ in the external gluon field of transverse size 
$\lambda_\perp \sim 1/k_{2t}$.
However, in the kinematical region \eqref{eq:wrong} the transverse size $1/k_{1t}$ of the
parton fluctuation $p\to p'+k_1$ is {\em smaller}\/ than the resolution power of the probe,
$\lambda_\perp$. In these circumstances the destructive interference between $k_2$ interacting 
with the initial ($p$) and final state ($p'+k_1$) comes onto the stage and eliminates 
\eqref{eq:wrong} reducing lifetime ordering to the transverse momentum one, 
\beq
\big( \gone + R \big) - V\>: \qquad  {\bf  k}_t^2/k_+ \> \Longrightarrow \>   {\bf k}_t^2. \nonumber
\eeq
In the language of the RRE, it is the r\^ole of generated DL terms to take care of 
this transformation due to coherent suppression of the part of
the phase space \eqref{eq:wrong}  kinematically allowed to soft gluons by 
the dynamically unaware, unsuspecting lifetime ordering.

For the time-like evolution,  the very same generated double logs 
 {\em add}\/ to those in $\cP$, resulting in the characteristic DL series for the
time-like anomalous dimension,
\beq
  \gamma^{(\mbox{\scriptsize T})}_{\mbox{\scriptsize DL}}  
  = \quart \bigg( \sqrt{\omega^2 + 8\alphys} -\omega \bigg)
  = \frac{\alphys}{\omega} - 2 \frac{\alphys^2}{\omega^3} + 8\frac{\alphys^3}{\omega^5} -40\frac{\alphys^4}{\omega^7} + 224\frac{\alphys^5}{\omega^9} +  \ldots \,,
\eeq
whose first three terms are present in \eqref{PsmallT}. 
This structure of the time-like anomalous dimension  is equivalent to ordering 
soft gluon cascades in {\em angles}:
\beq
\big( \gone + R \big) + V\>: \qquad   {\bf k}_t^2/k_+ \>\to\>   {\bf k}_t^2 /k_+^2.\nonumber
\eeq
The angular ordering following, once again, from destructive soft
gluon interference in the angle disordered kinematics~\cite{AO}, can be considered 
to be an image of the transverse momentum ordering in the space-like case, via the RRE. 

In spite of the fact that the evolution kernel $\gone$ does not
correspond to a clever choice of the evolution variable in either T-
or S- channel, its universality can be exploited to relate DIS and $e^+e^-$ anomalous
dimensions. In particular, the RRE links together two puzzling results that were never
thought to be of a common origin \cite{DMSunder}.  
They are: the absence of the $\as^2$ and $\as^3$ terms in the BFKL anomalous dimension 
\eqref{PsmallS} in the DIS problem  
and, on the other hand,  phenomenon of {\em exact}\/ angular ordering \cite{EAO,MLLA}
that seems to hold, unexpectedly, 
down to the next-to-next-to-leading order in $\ee$ anomalous dimension \cite{Malaza}.

\section{Conclusions \label{Sec-Conc}}

Maximal transcedentality Euler--Zagier harmonic sums that describe anomalous 
dimensions of twist-two operators in ${{\cal N}}\!=\!4$ super-symmetric Yang--Mills theory 
are genetically related, from QFT point of view, to multiple radiation of soft gluons. 
The word ``soft'' here stands for lack of  better terminology. In fact, these gluons may
be perfectly ``hard'', carrying a large (light-cone) momentum fraction $x=\cO{1}$.  
One would rather call them {\em classical}. The point is as follows. 
Mellin image of the Born level  ${{\cal N}}\!=\!4$ SYM anomalous dimension, 
\beq\label{gLBK}
\gamma_1(N) =-S_1(N) = \psi(N+1)-\psi(1) \quad \Longrightarrow\quad
 \tilde{\gamma}_1(x) \>=\> 
 \frac{x}{1-x}  \quad (x<1),
\eeq
is nothing but the universal part of the gluon radiation probability which, according to the
celebrated Low--Burnett--Kroll theorem \cite{LBK} has {\em classical nature}. Such radiation
does not depend on quantum properties of the emitter, but its (colour) charge, and does not
affect quantum state of the radiating system. 
{\em Classical}\/ (or LBK) gluons  \eqref{gLBK} constitute a significant part
of diagonal QCD splitting functions,
\bsub
\beeq
  \tilde{\gamma}_{q\to q}(x) \>&=&\> \frac{C_F\as}{\pi}\left[\, \frac{x}{1-x} \>+ \frac{(1-x)}{2} \right], \\
   \tilde{\gamma}_{g\to g}(x) \>&=&\> \frac{C_A\as}{\pi} 
   \left[ \frac{x}{1-x} \>+ (1-x)\cdot \big(x+x^{-1}\big) \right]. 
\eeeq
\esub
The {\em quantum}\/ parts of emission probabilities, depending on the details of the emitter, 
are down by {\em two powers}\/ of the gluon momentum fraction $(1-x)$, in perfect accord with the
LBK wisdom.  

``Classical'' does not necessarily means ``simple''. However, amazing simplifications do arise, 
now and then, in the QFT context where classical radiation is concerned.  
Thus, {\em exact}\/ Parke--Taylor QCD amplitudes \cite{PT} 
which describe radiation of arbitrary number of gluons
with ``classical'' helicities,  $2\to n$, and generalise  production probabilities \cite{BCM}
of soft gluons with $x_i\ll 1$  to arbitrary momentum fractions  $x\sim 1$.
Hence, a breathtaking possibility of an {\em all-order}\/ prediction  \cite{ES} of the magnitude of the 
``cusp anomalous dimension''  \cite{cusp},
which is just another name for intensity of soft (classical) gluon radiation 
--- the ``physical'' (or ``bremsstrahlung'') QCD coupling \cite{CMW,DKT}. 

From this perspective, dynamics of the ${{\cal N}}\!=\!4$ SYM model has a good chance to be
``classical'', to extent, since,  thanks to super-symmetry, quantum effects due to fermions 
and bosons have cancelled both in the anomalous dimension 
 \eqref{gLBK} and in the  $\beta$-function. 
 This idea finds a strong support  in the conjectured gauge--string duality \cite{Polyakov} 
---  AdS/CFT correspondence \cite{AdS/CFT}. 
A powerful property of integrability,  both in the high energy 
Regge asymptotics of scattering amplitudes \cite{Lip93,FadKor} and in the scale dependence 
of quasi-partonic operators \cite{quasi-part} manifests itself most profoundly 
in the ${{\cal N}}\!=\!4$ SYM case \cite{integr}. 
Descending from SYM back to QCD by reducing the number of super-symmetries, one finds 
integrability property to survive in certain sub-sectors of QCD parton dynamics, such as  
baryon wave function and maximal helicity multi-gluon operators \cite{Braun}. 

%
All that makes this model a perfect playing ground for unveiling internal structure of multi-gluon 
radiation in QCD, whose effects spoil perturbative expansions and bear major 
responsibility for complexity 
of higher order calculations, and of QCD anomalous dimensions in particular. 
In the present work we chose ${{\cal N}}\!=\!4$ SYM model in order to learn how 
to single out higher loop effects, and structures, that are {\em inherited}\/ from 
the lower orders of the perturbative expansion. 
To this end we used bookkeeping based on the concept of the Gribov--Lipatov reciprocity 
respecting evolution equation, RRE, \cite{DMS,DKT} which treats space- and time-like parton 
multiplication in one go.

Power of the LBK wisdom about the classical nature of radiation described by \eqref{gLBK} 
revealed itself most remarkably in this model. 
The RRE applied to the space-like anomalous dimension $\guni$,  
known in three loops \cite{KLOV},  
has generated the major part of higher order contributions in terms of a compact
``evolution kernel'' $\gone$ linked to $\guni$ by a non-linear relation  \eqref{RRE}.
Simplicity of the evolution kernel is suggestive and invites deeper studies.

\paragraph{Note added.}
When this project was under completion, we learned about the work by 
B.~Basso and G.\ Korchemsky in which the reciprocity respecting equation \eqref{RRE} 
has naturally emerged as a consequence of the conformal symmerty.

\section*{Acknowledgements}

We are grateful to Gregory Korchemsky, Anatoly Kotikov, Lev Lipatov and Gavin Salam  
for illuminating discussions. 

\appendix

\setcounter{equation}{0}

\section{Harmonic sum representation \label{Aharm}}

The answer contains  {\em harmonic sums}, 
$$
 S_a = \sum_{k=1}^N   k^{-a}, \>\> 
 S_{-a} = \sum_{k=1}^N (-k)^{-a}; \quad  \hS{p}(N) \equiv S_{p}(N) - S_p(\infty),
$$ 
having the following Mellin representations: 
\bsub\label{SMel}
\beq
 S_1(N) = \>   \M{\frac{x}{(x-1)_+}}  = \psi(N+1) - \psi(1) \,,
\eeq 
and ($ a\ge 2$) 
\beeq
  S_{a}(N) - \zeta_a  \equiv \>\> \hS{a}(N) \> &=& \frac{(-1)^{a-1}}{\Gamma(a)}
\> \M{ \frac{ x \ln^{a-1} x}{x-1}  }    ; \\
  \label{Sneg}
 S_{-a}(N) + \zeta_a(1\!-\!2^{-a}) \equiv \hS{-a}(N)  &=&  \frac{(-1)^{a-1}}{\Gamma(a)} 
\> \M{  \frac{x \ln^{a-1} x}{x+1}} \cdot (-1)^N .
\eeeq 
\esub
As for generalized multi-index Euler--Zagier harmonic sums, a representation alternative to that in \eqref{eq:uni} is better suited for analysis of the large-$N$ asymptotics. 
It employs the basis of sums with the negative index moved from the head to the tail 
of the index vector,  $S_{\vec{m},-p}$ instead of $S_{-p,\vec{m}}$. The latter may contain a logarithmic
enhancement factor, $S_{-p,\vec{m}}\propto S_1^\ell(N)$, with $\ell$ the number 
of the rightmost  $+1$ indices, e.g., $S_{-2,1,1,1} \propto S_1^3 \hS{-2} = \cO{\ln^3N/N^2}$.  

Crucial simplification of the answer is achieved in terms of  {\em complementary harmonic sums}\/ 
$\hhS{\vec{m},-p}$
which we define by a recursive relation
\beq
 \hhS{a_1,\ldots,a_n}(N) \>=\> S_{a_1,\ldots,a_n}(N) 
 - \sum_{k=1}^{n-1}  S_{a_1, \ldots, a_k}(N) \cdot \hhS{a_{k+1},\ldots, a_n}(\infty), \quad n\ge 2.
\eeq
In particular,
\beq
 \begin{split}
  \hhS{a,-p}(N)  &=   \Ss(a,-b)(N) -  S_a(N)S_{-p}(\infty) , \\ 
 \hhS{a,b,-p}(N) &=  \Sss(a,b,-p)(N)  - \S(a)(N) \hhS{b,-p}(\infty) -  \Ss(a,b)(N)S_{-p}(\infty) , \\
 \hhS{a,b,c,-p}(N) &=  \Sss(a,b,c,-p)(N)  - \S(a)(N) \hhS{b,c,-p}(\infty) -  \Ss(a,b)(N)\hhS{c,-p}(\infty)
 -  \Ss(a,b,c)(N)S_{-p}(\infty)  . \nonumber
\end{split}
\eeq
These functions are finite at $N=\infty$. 
Subtracting the value at infinity, one arrives at \eqref{Scompl}.
The case of all positive indices $a_i$ but the very last one, $a_n=-p$,
 is special in that such sums fall fast with $N$ and contain no $\ln N$ factors accompanying
 {\em  any subleading power}\/ $N^{-k}$. 

Mellin transforms of participating multi-index sums are as follows:
\bsub
\beeq\label{negsum}
 \Gamma  \hhhS{a,-p} &= &  \M{ \frac{x}{1+x}  \int_x^1 \frac{dz}{z+1}\, 
{\ln^{a-1}\frac{z}{x}}\, {\ln^{p-1} \frac1z } } \nonumber \\
 \Gamma \hhhS{a,b,-p} &=&  \M{\frac{x}{1+x}\,  \int_x^1  \frac{dz}{1+z} \ln^{p-1}\frac1{z} 
 \int_x^z \frac{dt}{1+t} \ln^{b-1}\frac{z}{t} \, \ln^{a-1}\frac{t}{x}} ,\\
 \Gamma \hhhS{a,b,c,-p} &=&   \M{ \frac{x}{1\!+\!x}  \int_x^1 \frac{dy}{1\!+\!y}  \ln^{a-1}\frac{y}{x}
   \int_y^1 \frac{dt}{1\!+\!t} \, \ln^{b-1}\frac{t}{y}  \int_t^1  \frac{dz}{1\!+\!z}
\ln^{c-1}\frac{z}{t} \> \ln^{p-1}\frac1{z} },\nonumber 
\eeeq
where
\beq
 \Gamma \>=\> (-1)^N \cdot \bigg( \prod_{g\in (a, \ldots, p )} \Gamma(g) \bigg)^{-1} .
\eeq
\esub
Constructing linear combinations \eqref{Ys} using \eqref{negsum}, one arrives at \eqref{Ydef}.

\section{Shifting negative indices to the right}

Here we list relations that help to transform the original representation \eqref{eq:uni} 
in terms of harmonic sums
$S_{-p,\vec{m}}$ of \cite{KLOV} into the negative-on-the-right form, $S_{\vec{m},-p}$:

\noindent
Two indices
\begin{eqnarray*}
 \S(-2,1) &=& -\S(1,-2) + S_1\S(-2) + \S(-3), \quad  \S(-2,2) = -\S(2,-2) + S_2\S(-2) + \S(-4), \\
 \S(-2,3) &=& -\S(3,-2) + S_3\S(-2) + \S(-5); \quad  \S(-3,1) = -\S(1,-3) + S_1\S(-3) + \S(-4), \\
 \S(-3,2) &=& -\S(2,-3) + S_2\S(-3) + \S(-5); \quad  \S(-4,1) = -\S(1,-4) + S_1\S(-4) + \S(-5).
\end{eqnarray*}
Three indices
\begin{eqnarray*}
 S_{-2,1,1} &=& S_{1,1,-2} + S_1\big(S_{-3}-S_{1,-2} \big) + S_{-2}S_{1,1}
 - S_{2,-2} - S_{1,-3} + S_{-4} , \qquad { } \\
 S_{-3,1,1} &=&  S_{1,1,-3} + S_1\big(S_{-4}-S_{1,-3} \big) + S_{-3}S_{1,1}
 - S_{2,-3} - S_{1,-4} + S_{-5} ; \\
 S_{-2,2,1} &+& S_{-2,1,2} =  (S_{2,1,-2} + S_{1,2,-2}) + S_1(S_2S_{-2}-S_{2,-2} + S_{-4})   \\
 && \quad +\> S_2(S_{-3} -S_{1,-2}) + S_3S_{-2} -2S_{3,-2} -S_{2,-3}  -S_{1,-4} +2S_{-5} .
\end{eqnarray*}
Four indices
\beq\begin{split}
3S_{-2, 1, 1, 1} &= -3S_{1, 1, 1, -2}  + S_{1,1}(2S_{-3}-3S_{1, -2}) + S_{1, 1}S_{-3}  
+  S_{3}S_{-2}   \nonumber\\
&+S_{1}  \big(3S_{1, 1, -2} +(S_{1, 1}+S_2)S_{-2} -3S_{2, -2} -3S_{1, -3} +3S_{-4} 
\big)\\
&+  3(S_{1, 2, -2} + S_{2, 1, -2} +S_{1, 1, -3}) 
 - 3(S_{1, -4} +S_{2, -3} +S_{3, -2}  -S_{-5}) .
\end{split}\eeq


\end{document}